\documentclass[12pt]{iopart}

\usepackage{iopams}
\usepackage{setstack}
\usepackage{graphicx}
\usepackage{color}
\usepackage{epsfig}
\usepackage{psfig}
\usepackage{pstricks}

\usepackage{setspace}
\usepackage[breaklinks=true, debug=true, colorlinks=true, ps2pdf]{hyperref}

\makeatletter

\begin{document}

\title{$J/\Psi$ production in $Cu+Cu$ and $Au+Au$ collisions
measured by PHENIX at RHIC}

\author{Andry M. Rakotozafindrabe, for the PHENIX collaboration}

\address{Laboratoire Leprince Ringuet, \'{E}cole Polytechnique, Route de Saclay, 91128 Palaiseau, France}
\begin{abstract}
PHENIX preliminary results on the $J/\Psi$ production in $Cu+Cu$ and $Au+Au$ collisions at $\sqrt{s_{NN}}=200$~GeV are presented. They are compared to results from lower energy experiments NA50 and NA60 at CERN SPS and to expectations from various theoretical models.
\end{abstract}

\submitto{\jpg}

\section{Introduction}

Based on the assumption that $c\bar{c}$ pairs are created in primordial hard nucleon-nucleon collisions only (due to the large mass of the charm quark), $J/\Psi$'s suppression by color screening in a deconfined medium~\cite{Matsui_Satz} is suggested to be used as a signature of a Quark-Gluon Plasma (QGP) formation in the relativistic heavy ion collisions. However, at RHIC energies, up to $N_{c\bar{c}}\approx10$-20~$c\bar{c}$ pairs can be created in the most central $Au+Au$ collisions. Since the probability to recombine uncorrelated $c$ and $\bar{c}$ goes as $N_{c\bar{c}}$, in-medium formation of $J/\Psi$'s is foreseen at RHIC, leading to an enhancement of the $J/\Psi$ yield~\cite{Grandchamp,Thews}. Alternate mechanisms, not involving QGP, such as the interaction with hadronic co-movers~\cite{Capella} (which are secondaries produced in the collision), also successfully accounted for the abnormal $J/\Psi$ suppression in most central $Pb+Pb$ collisions, in excess of the normal nuclear absorption, as seen by the lower energy experiment NA50~\cite{NA50_2000,NA50_2005} at CERN SPS. The measurements at higher energy could help disentangling this variety of models. In addition to these effects which can be expected in case of the QGP formation, other so-called cold nuclear effects~\cite{Vogt} are known to affect the $J/\Psi$ yield, in particular: \emph{(i)} the shadowing (initial-state effect) which is a modification of the parton distribution function of a free nucleon by the nuclear environment, and \emph{(ii)} the nuclear absorption (final-state effect) where the formed $c\bar{c}$ pre-resonance is broken by interactions with primary target/projectile nucleons. $d+Au$ measurements is used to evaluate the cold nuclear effects at RHIC. 

\begin{figure}[htb]
\vspace*{-2mm}
   \begin{minipage}[c]{.5\linewidth}
   		\begin{center} 
				\includegraphics[width=0.9\linewidth, height=6.5cm]{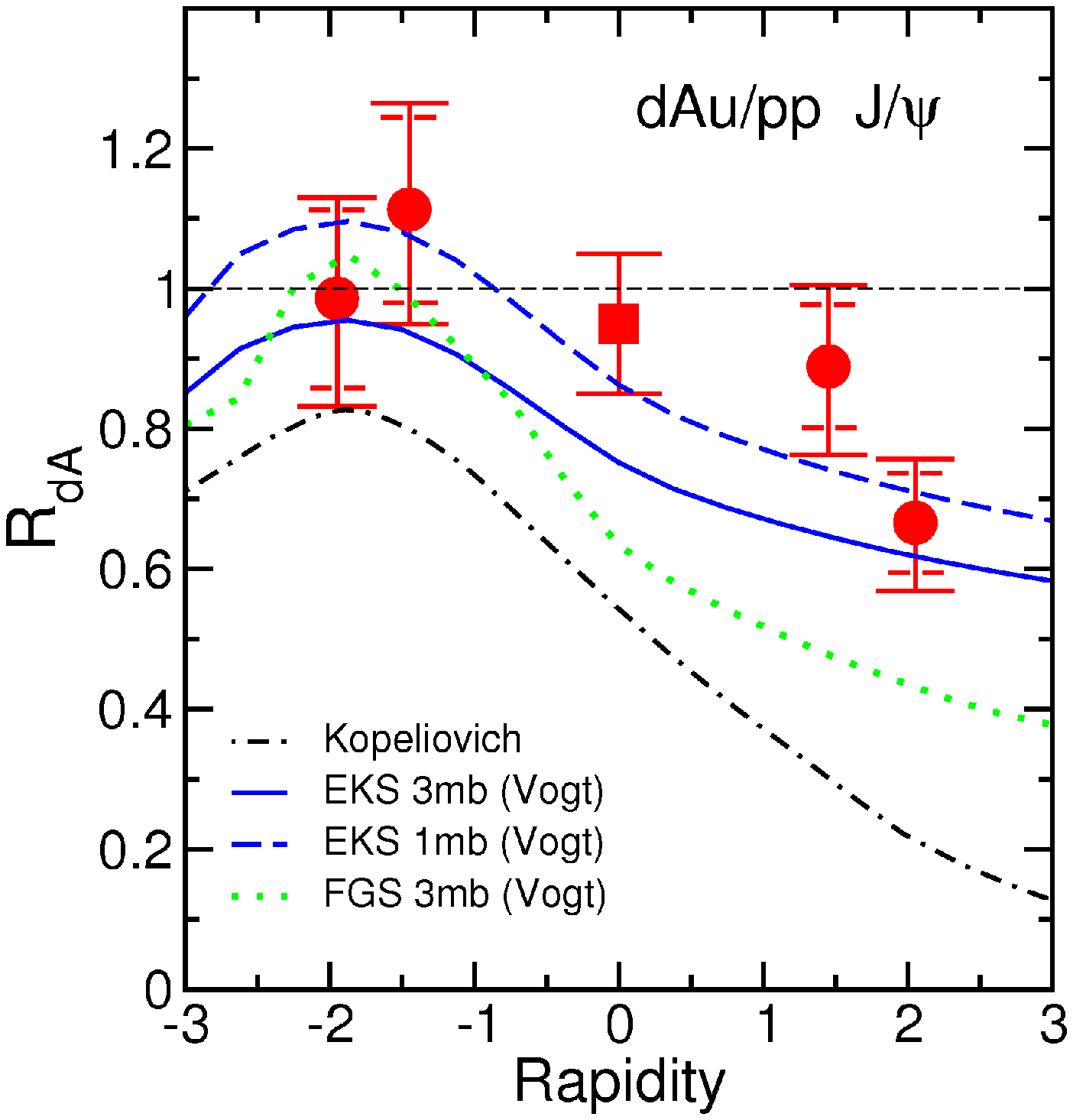}
    	\end{center}
   \end{minipage} \hfill
   \begin{minipage}[c]{.5\linewidth}
   	 \begin{center} 
				\includegraphics[width=0.9\linewidth, height=6.5cm]{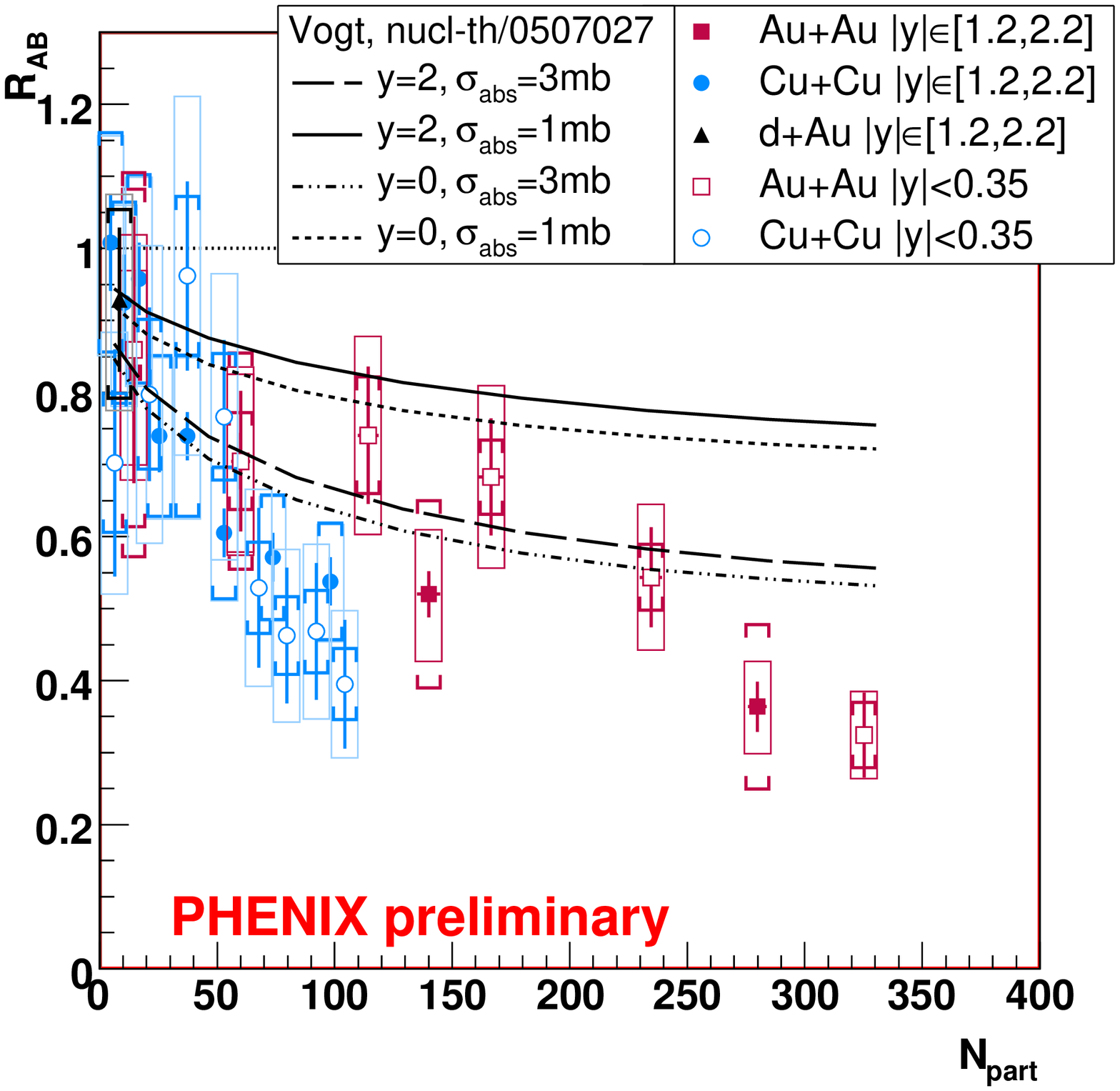}
     \end{center}
   \end{minipage}
   \vspace*{-6mm}
 \caption{Left panel: $J/\Psi$ nuclear modification factor measured in $d+Au$ as a function of rapidity and compared to theoretical expectations for cold nuclear effects~\cite{dAu_pp_results}. Kopeliovich~\cite{Kopeliovich} and FGS parametrization use a stronger gluon shadowing than the EKS98 parametrization~\cite{EKS98_FGS_2003,EKS98_FGS_2005}. Right panel: $J/\Psi$ nuclear modification factor measured in $d+Au$, $Cu+Cu$ and $Au+Au$ vs the number of participating nucleons~\cite{Hugo} compared to Vogt~\cite{Vogt} predictions in $Au+Au$ collisions for cold nuclear effects. Vertical bars stand for statistical errors, brackets for point-to-point and boxes for global systematic errors.}
 \label{fig:cold_eff}
 \vspace*{-3mm}
\end{figure}

\section{Results and discussion}

\begin{figure}[htb]
\vspace*{-2mm}
   \begin{minipage}[c]{.315\linewidth}
   		\begin{center} 
				\includegraphics[width=\linewidth]{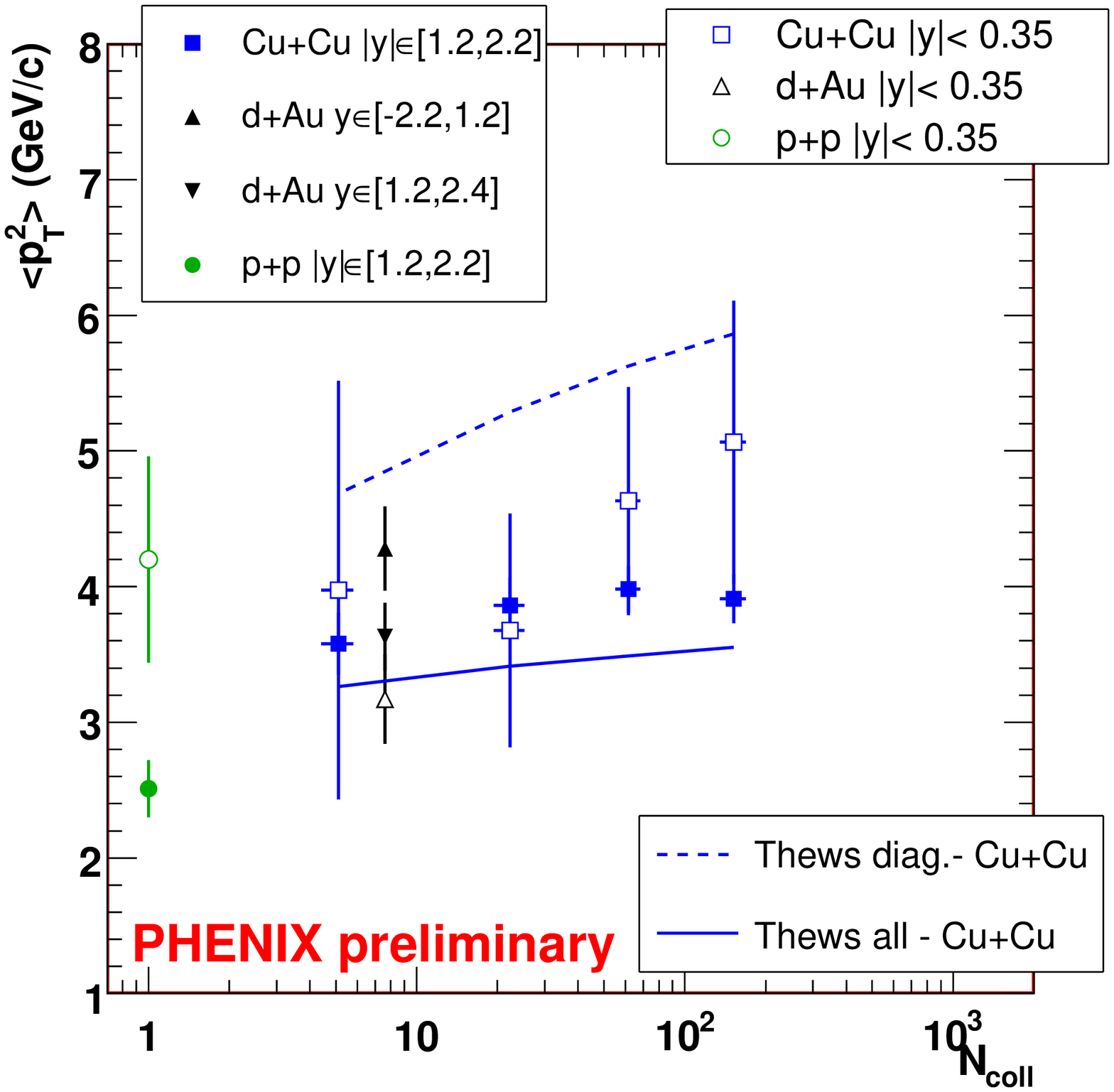}
    	\end{center}
   \end{minipage} \hfill
   \begin{minipage}[c]{.315\linewidth}
   	 \begin{center} 
				\includegraphics[width=\linewidth]{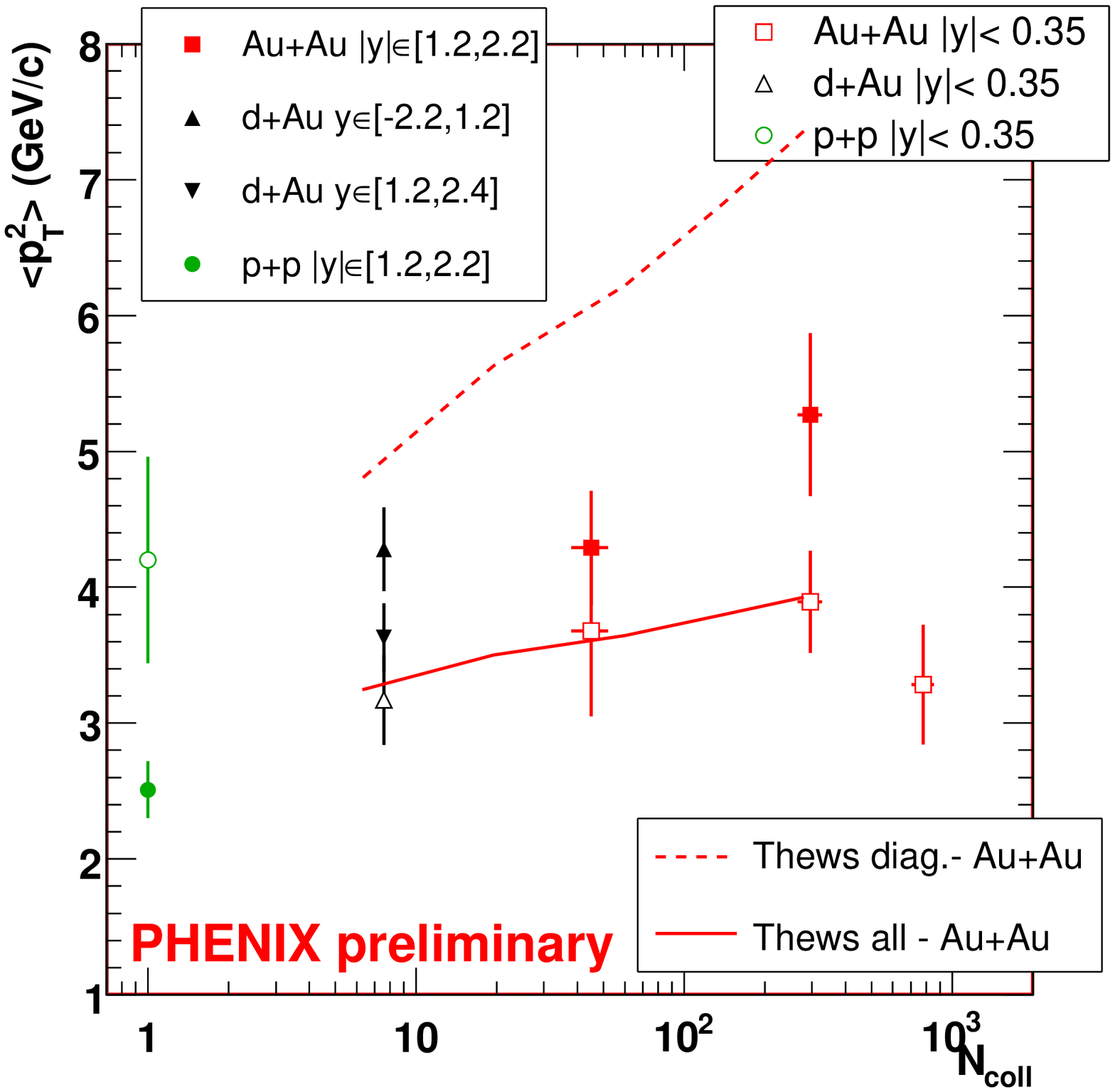}
     \end{center}
   \end{minipage}
      \begin{minipage}[c]{.35\linewidth}
   	 \begin{center} 
				\includegraphics[width=\linewidth]{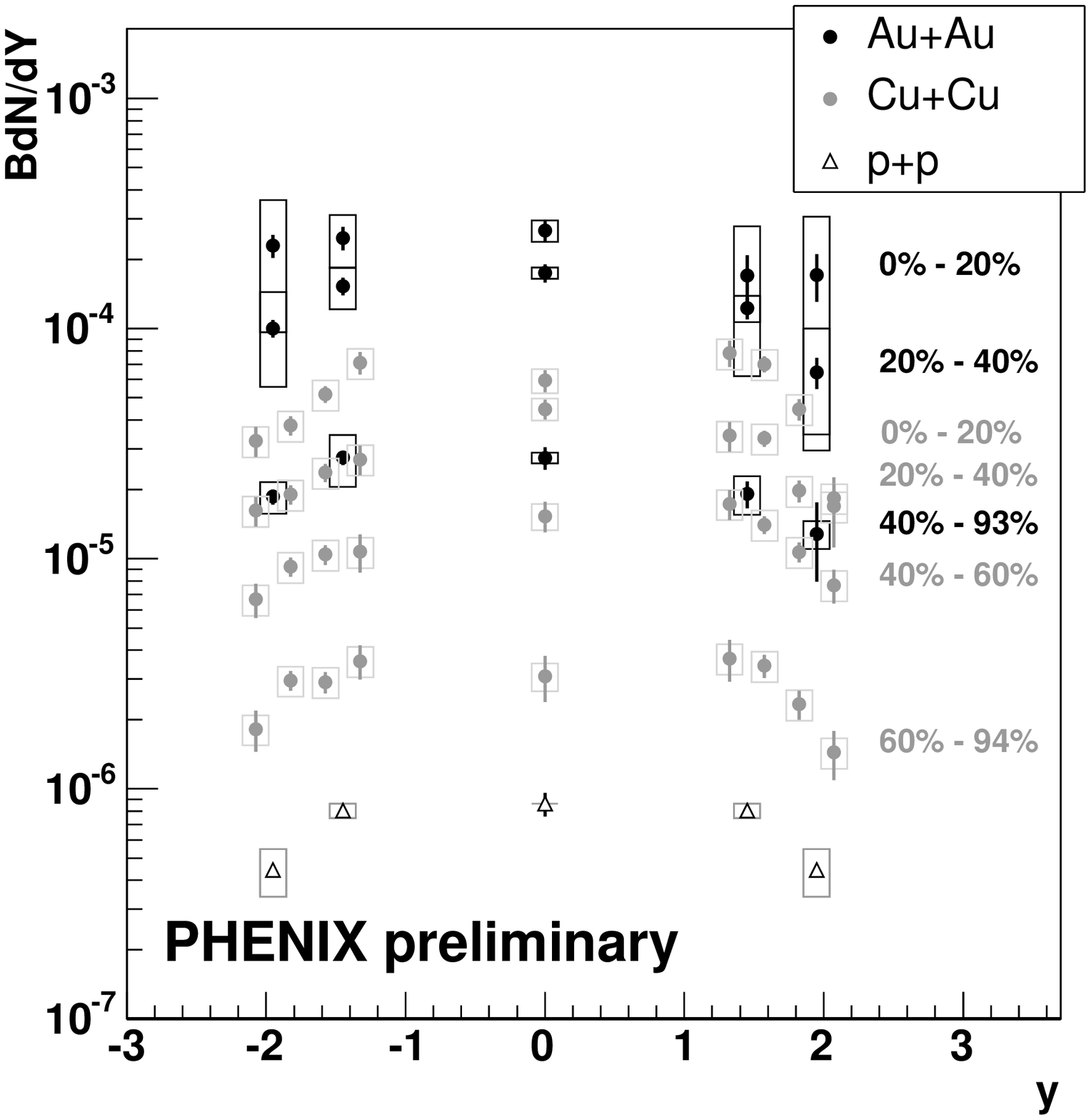}
     \end{center}
   \end{minipage}
   \vspace*{-5mm}
 \caption{$J/\Psi$ mean square transverse momentum as a function the number of binary collisions in $p+p$, $d+Au$, $Cu+Cu$ (left panel) and $Au+Au$~\cite{Hugo} (middle panel). It is compared to Thews predictions~\cite{Thews} with recombination (solid line) and without recombination (dashed line). Open makers are for the mid rapidity measurements, and solid markers for the forward rapidity ones. Right panel: $J/\Psi$ yield as a function of rapidity and increasing centrality from $p+p$ to $Cu+Cu$ to most central $Au+Au$~\cite{Hugo}. Vertical bars stand for statistical errors and boxes for point-to-point systematic errors.}
 \label{fig:test_recombination}
 \vspace*{-6mm}
\end{figure}

During the RHIC data taking periods Run-3 to Run-5 (2003-2005) PHENIX measured the $J/\Psi$ yield in $p+p$, $d+Au$, $Cu+Cu$ and $Au+Au$ at $\sqrt{s_{NN}}=200$~GeV. $J/\Psi$ measurements are performed in two decay channels, $J/\Psi\rightarrow\e^+~e^-$ at mid rapidity ($|y|<0.35$), and $J/\Psi\rightarrow\mu^+~\mu^-$ at forward rapidity ($1.2<|y|<2.2$) intervals. Here, we will focus on $Cu+Cu$ and $Au+Au$ preliminary results (resp. Run-5 and Run-4). Run-3 $p+p$ results~\cite{dAu_pp_results} define the baseline of the $J/\Psi$ production and hence are used to compute the nuclear modification factor defined as $R_{AB}=\frac{dN^{AB}/dy}{N_{coll}^{AB}dN^{pp}/dy}$ where $N_{coll}$ is the number of binary collisions. $R_{AB}$ is expected to be unity if there is no effect. The Run-3 $d+Au$ results~\cite{dAu_pp_results} are used to characterize the cold nuclear effects at RHIC. 
The left panel of figure~\ref{fig:cold_eff} shows the nuclear modification factor $R_{dAu}$ as a function of rapidity. At positive rapidity ($d$-going direction) are probed $Au$ partons carrying a small momemtum fraction $x$, which is the region where the gluon shadowing is expected. $R_{dAu}$ is indeed significantly lower than unity at positive rapidity. Theoretical  expectations~\cite{Kopeliovich,EKS98_FGS_2003,EKS98_FGS_2005} for nuclear absorption and shadowing effects are shown for comparison. Data favors a modest shadowing in agreement with the EKS98 parametrization and a modest absorption cross-section $\sigma_{abs}=1$~mb, but a stronger absorption ($3$~mb) is still compatible within our large errors. A higher luminosity $d+Au$ data set is desirable to better constrain the cold nuclear effects at RHIC. 
The preliminary results for the centrality dependence of the nuclear modification factor in $Cu+Cu$ and $Au+Au$~\cite{Hugo} are shown on the right panel of figure~\ref{fig:cold_eff}, together with the published minimum bias $d+Au$ value. The suppression pattern are compared to the predictions~\cite{Vogt} which use EKS98 shadowing parametrization for central ($y=0$) and forward rapidities ($y=2$) and assume $1$~mb and $3$~mb nuclear absorption. Even relative to the $3$~mb nuclear absorption case, the $J/\Psi$ suppression in most central $Au+Au$ collisions goes significantly beyond the cold nuclear effects at RHIC. 
A factor $\approx3$ of suppression is seen relative to binary scaled~$p+p$. This is surprisingly close to the suppression factor seen at lower $\sqrt{s_{NN}}=17.3$~GeV by the NA50 experiment~\cite{NA50_2000,NA50_2005} for the most central $Pb+Pb$ collisions  relative to the NA51 $p+p$ result~\cite{NA51}. Moreover, suppression at RHIC energy follows the same general trend~\cite{ViNham} as NA50 (and consequently preliminary NA60~\cite{NA60}) suppression, in spite of very different conditions, namely $\sqrt{s_{NN}}$, cold nuclear effects (at SPS, $\sigma_{abs}=4.18\pm0.35$~mb), and the estimated maximum energy density ($\approx3$~GeV/fm$^3$ at SPS, $5$~GeV/fm$^3$ at RHIC, if assuming the same formation time of $1$~fm$/c$). 
\begin{figure}[tbh]
\vspace*{-2.5mm}
   \begin{minipage}[c]{.5\linewidth}
   		\begin{center} 
				\includegraphics[width=0.9\linewidth, height=6.5cm]{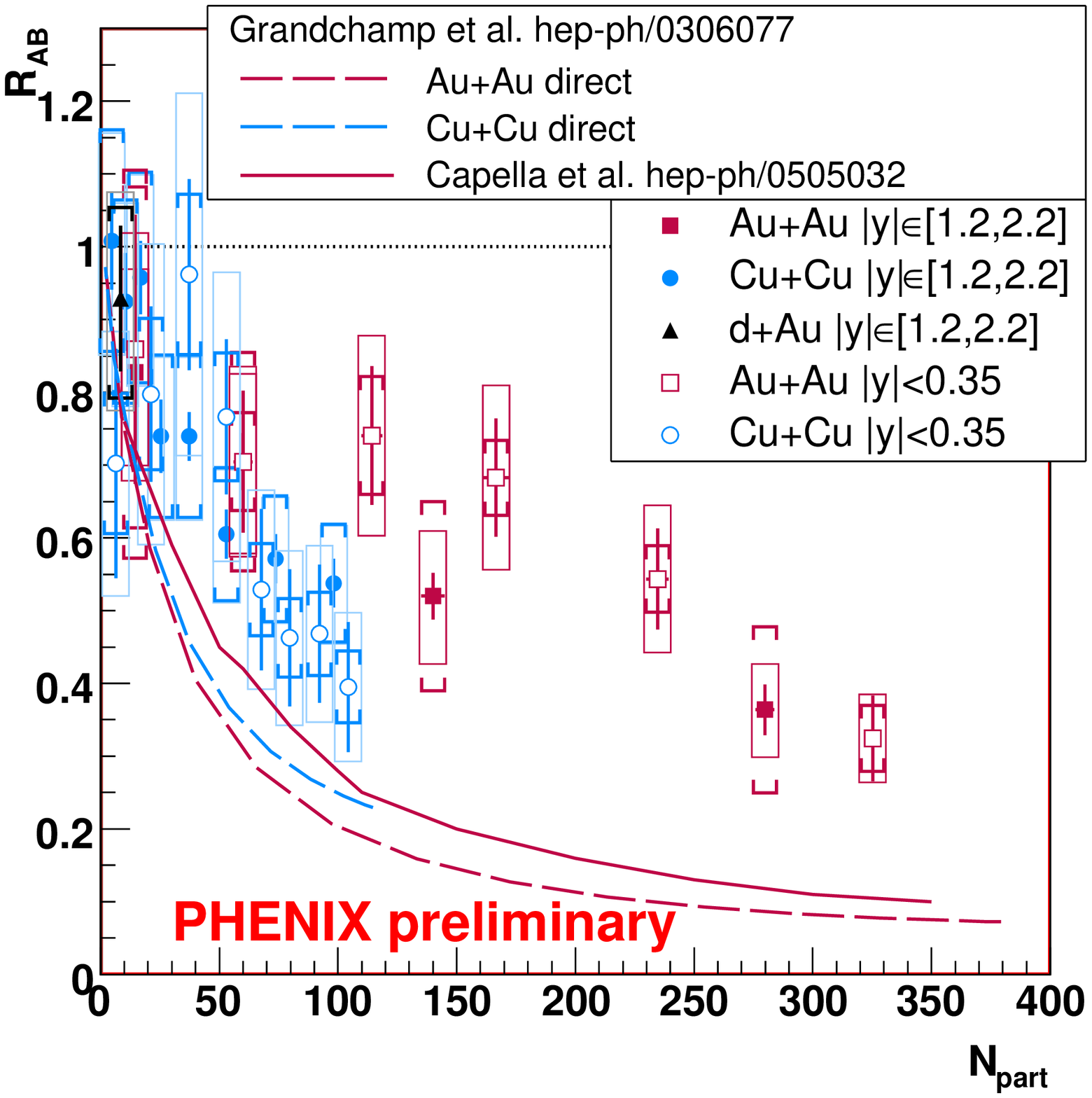}
    	\end{center}
   \end{minipage} \hfill
   \begin{minipage}[c]{.5\linewidth}
   	 \begin{center} 
				\includegraphics[width=0.9\linewidth, height=6.5cm]{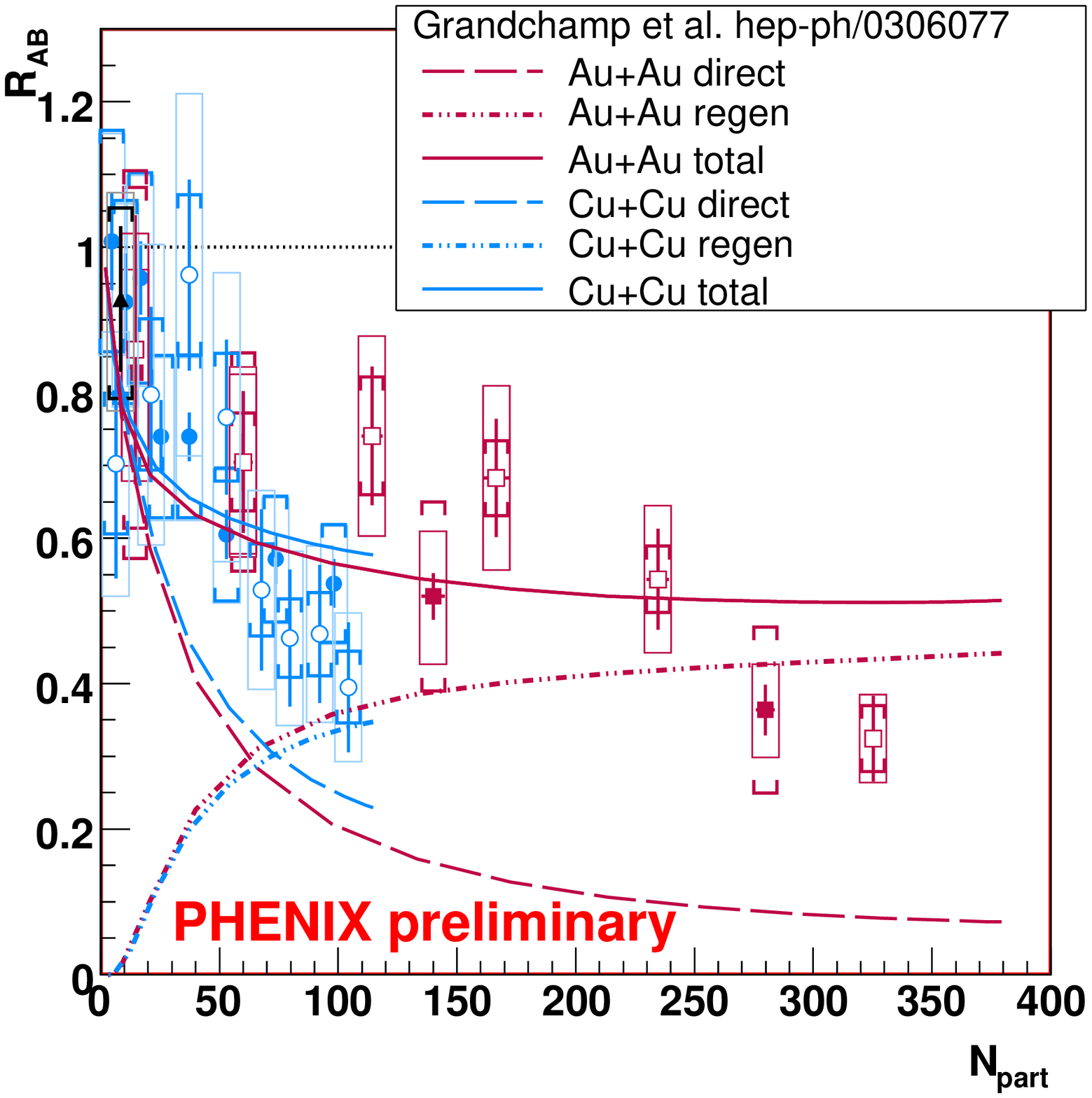}
     \end{center}
   \end{minipage}
   \vspace*{-5mm}
 \caption{$J/\Psi$ nuclear modification factor in $d+Au$, $Cu+Cu$ and $Au+Au$ vs the number of participating nucleons~\cite{Hugo}  compared to various models of final-state interaction in the medium. Left panel: solid line stands for the co-mover model~\cite{Capella}, dashed lines for a direct suppression in a hot medium~\cite{Grandchamp}. Right panel: solid lines is the sum of the direct suppression (dashed lines) and the regeneration~\cite{Grandchamp} (dotted lines).}
 \label{fig:models_vs_data}
 \vspace*{-5mm}
\end{figure}
On the left panel of figure~\ref{fig:models_vs_data}, the centrality dependence of the nuclear modification factor is compared to two different models of final-state interactions that successfully described the NA50 results: the first one suppresses $J/\Psi$'s using co-movers interactions~\cite{Capella}, and the second involves direct dissociation in a hot medium~\cite{Grandchamp}. Both models seem to overestimate the $J/\Psi$ suppression at RHIC energy. 
On the right panel of figure~\ref{fig:models_vs_data} the charm quark recombination mechanism~\cite{Grandchamp} is added to the direct dissociation and the sum is in better agreement with the observed suppression. 
To explore further the recombination scenario, more observables are compared to the related predictions from \cite{Thews} on figure~\ref{fig:test_recombination}, namely the centrality dependence of the mean square transverse momentum $<p_T^2>$ (left and middle panel) and the evolution with centrality of the $J/\Psi$ yield as a function of rapidity (right panel). In this model, recombined $J/\Psi$ are predicted to have a lower $<p_T^2>$ since charm quark production is supposed to be significant at low $p_T$. Observed values of $<p_T^2>$ have large errors and lay between the two hypotheses, with and without recombination. $J/\Psi$'s from recombination should also have a rapidity spectra peaked at $y=0$ where the charm density is expected to be larger. This implies a narrowing rapidity spectra with an increasing centrality, which seems not being observed here.

\section{Summary}
PHENIX preliminary results on $J/\Psi$ production measured at mid and forward rapidities in $Cu+Cu$ and $Au+Au$ at $\sqrt{s_{NN}}=200$~GeV show a suppression pattern well beyond cold nuclear effects, but still similar to the suppression seen at CERN SPS. Models that successfully describe SPS results overestimate the suppression at RHIC. Adding recombination seems to restore the agreement for the centrality dependence, but conclusions from testing recombination against over observables (mean square transverse momentum, shape of the rapidity spectra) remain unclear.





\end{document}